\documentclass[pra,twocolumn,aps,showpacs,superscriptaddress]{revtex4}

\usepackage{amsmath}
\usepackage{amssymb}
\usepackage{hyperref}
\usepackage{graphicx}
\usepackage{subfigure}
\usepackage[usenames,dvipsnames]{color}
\usepackage{setspace}

\begin{document}


\title{Anisotropic merging and splitting of dipolar Bose-Einstein condensates}

\author{S. Gautam}
\affiliation{Department of Physics, Indian Institute of Science,
             Bangalore - 560 012, India}
\author{Subroto Mukerjee}
\affiliation{Department of Physics, Indian Institute of Science,
             Bangalore - 560 012, India}


\date{\today}
\begin{abstract}
 We study the merging and splitting of quasi-two-dimensional
Bose-Einstein condensates with strong dipolar interactions. We observe
that if the dipoles have a non-zero component in the plane of the condensate,
the dynamics of merging or splitting along two orthogonal directions,
parallel and perpendicular to the projection of dipoles on the plane of the 
condensate are different. The anisotropic merging and splitting of the 
condensate is a manifestation of the anisotropy of the roton-like mode in the 
dipolar system. The difference in dynamics disappears if the dipoles are 
oriented at right angles to the plane of the condensate as in this case the 
Bogoliubov dispersion, despite having roton-like features, is isotropic.
\end{abstract}


\maketitle


\section{Introduction}
\label{I}
The first experimental generation of a Bose-Einstein condensate (BEC) in a
gas of chromium ($^{52}$Cr) atoms \cite{Griesmaier,Lahaye-1,Koch}, which have
permanent magnetic dipole moments, has lead to a flurry of experimental and
theoretical investigations on dipolar quantum gases. These have been
reviewed in Refs. \cite{Lahaye-2, Baranov}. Besides $^{52}$Cr, dipolar
Bose-Einstein condensates (DBECs) of dysprosium ($^{164}$Dy) \cite{Lu-1} and
erbium ($^{168}$Eb) \cite{Aikawa} have also been experimentally realized. 
Quantum degeneracy has also been realized in dipolar Fermi gases of 
dysprosium ($^{161}$Dy) \cite{Lu-2} and erbium ($^{167}$Eb) \cite{K_Aikawa}.
The two important characteristics of the dipolar interaction are its long 
range and anisotropic nature. The anisotropy introduced by the dipolar 
interactions has been observed in the the expansion dynamics \cite{Stuhler} 
and the collective excitations of a $^{52}$Cr condensate \cite{Bismut-2}. In 
contrast to contact interactions, all partial waves contribute to the 
scattering amplitude in the case of dipolar interactions. This makes the 
inter-particle interactions momentum dependent \cite{Lahaye-2}. As a 
consequence, the DBECs can support roton like excitations 
\cite{Santos,Wilson-1, Wilson-2}. The presence of a roton like mode in 
spectrum of the dipolar Bose-Einstein condensate (DBEC) lowers the Landau 
critical speed \cite{Landau} below the speed of sound for sufficiently large 
particle number\cite{Wilson-2}. 
Roton excitations can lead to density 
fluctuations at defects like vortices \cite{Wilson-1,Yi} and a roton 
instability \cite{Santos,Ronen,Wilson-3}. 
It has been demonstrated theoretically that the static and dynamic
structure factors which can be measured using Bragg spectroscopy
\cite{Blakie}, atom-number fluctuations in a trapped DBEC \cite{Bisset},
and the response of the condensate to weak lattice potentials 
\cite{Corson-1,Corson-2,Jona-Lasinio-1} can reveal the presence of the, still 
experimentally elusive, roton excitations. Roton excitations also enhance the 
density fluctuations in two-dimensional DBECs \cite{Boudjemaa}. The density 
dependence of the roton minimum also results in the spatial confinement of 
the rotons in trapped DBECs \cite{Jona-Lasinio-2}. In addition to these, 
dipolar interactions lead to anisotropic superfluidity, which, strictly 
speaking, is just the anisotropic manifestation of a roton like mode in the 
dipolar system \cite{Fischer,Ticknor,Muruganandam}. The anisotropic excitation
spectrum of the dipolar condensate of $^{52}$Cr has also been measured 
experimentally \cite{Bismut-1}. In the present work, we study the dynamics of 
the merging and splitting of the DBEC of $^{52}$Cr with a partially tilted 
polarization into the plane of the motion using a non-local Gross-Pitaevskii 
equation. The anisotropic coherence properties of such a DBEC at finite 
temperatures have been studied using Hartree-Fock-Bogoliubov method within 
the Popov approximation (HFBP) \cite{Ticknor-2}. We find that the anisotropic
superfluidity of these systems manifests itself as the directional dependence
of the merging and splitting dynamics.

The non-adiabatic merging and splitting of Bose-Einstein condensates (BECs) 
with pure contact interactions is known to lead to the formation of dispersive
shock waves \cite{Chang}. In the context of shock waves, the decay of small
density defects into quantum shock waves was earlier observed in BECs
\cite{Dutton}. The non-adiabatic collision of two strongly interacting
Fermi gases leading to the formation of shock waves has also been studied
\cite{Joseph,Bulgac,Ancilotto}. The propagation and nonlinear response
of dispersive shock waves, including the interaction of colliding shock waves,
in one and two-dimensional non-linear Kerr like media have also been
investigated \cite{Wan}. The interatomic interactions in these various
studies were isotropic. This is no longer the case for DBECs, which are the
focus of our study in the present work.

The paper is organized as follows- We start by providing the mean-field
description of DBECs in Sec.~\ref{Sec-II}. Here we discuss the 
quasi-two-dimensional non-local Gross-Pitaevskii equation (GPE) which we employ
to study the DBEC with an arbitrary direction of polarization.
In Sec.~\ref{Sec-III}, after analyzing the validity of the 
quasi-two-dimensional GPE, we numerically investigate the splitting and 
merging dynamics of DBEC. We finally conclude by providing a summary of 
results in Sec.~\ref{Sec-IV}.

\section{Mean Field Description of Dipolar Bose-Einstein condensate}
\label{Sec-II}
In the mean field regime, a scalar DBEC at $T = 0$ K can be well described by
the non-local GPE
\cite{Lahaye-2, Baranov}
\begin{eqnarray}
i\hbar\frac{\partial \Phi(\mathbf x,t)}{\partial t} &=&
\left[-\frac{\hbar^2\nabla^2}{2m}
+ V(\mathbf x) +g|\Phi(\mathbf x,t)|^2 \right.
\label{Eq.1}
\\
 &&\left.+\int U_{\rm dd}(\mathbf{x-x'})
|\Phi(\mathbf x',t)|^2 d\mathbf x'\right] \Phi(\mathbf x,t),\nonumber
\end{eqnarray}
where $\Phi(\mathbf x, t)$ is the wave function of the condensate. The
non-local term, the last term in the parenthesis, accounts for the long range
dipole-dipole interaction. In case the dipolar gas is polarized, i.e., all
the dipoles are oriented along the same direction, the dipole-dipole 
interaction energy is
\begin{equation}
U_{\rm dd} = \frac{C_{\rm dd}}{4\pi}
             \frac{1-3\cos^2\theta}{|\mathbf x-\mathbf x'|^3},
\end{equation}
where $\theta$ is the angle between the direction of polarization and relative
position vector of the dipoles. The coupling constant
$C_{\rm dd} = 12\pi\hbar^2a_{\rm dd}/m$, where $a_{\rm dd}$ is the length
characterizing the strength of the dipolar interactions 
and $m$ is the mass of the atom. For the dipolar gas
consisting of atoms with permanent magnetic dipole moment $\chi$ like
$^{52}$Cr, $a_{\rm dd} = \mu_0\chi^2m/(12\pi\hbar^2)$, where $\mu_0$ is the
permeability of the free space. The harmonic trapping potential
$V (\mathbf x) = m(\omega_x^2 x^2 + \omega_y^2 y^2 + \omega_z^2 z^2 )/2$,
where $\omega_j$'s with $j=x,y,z$ are trapping frequencies along the three 
coordinate axes. The contact interaction between atoms
is characterized by the interaction strength $g = 4\pi\hbar^2 a/m$, where $a$
is the $s$-wave scattering length. The total number of atoms $N$ and energy
$E$ are conserved by Eq.~(\ref{Eq.1}). For the sake of solving
Eq.~(\ref{Eq.1}) numerically, we transform the GP equation into
dimensionless form using following transformations:
\begin{eqnarray}
\mathbf x &= &\tilde{\mathbf {x}} a_{\rm osc},~t  =
2\tilde {t} \omega^{-1},~
\Phi = \sqrt{N}\tilde {\Phi}a_{\rm osc}^{-3/2},
\end{eqnarray}
where $a_{\rm osc} = \sqrt{\hbar/(m\omega)}$ with $\omega =
\rm min~ \{\omega_x,\omega_y,\omega_z\}$ is the oscillator length.
The dimensionless GP equation for the DBEC is now of the form
\begin{eqnarray}
i\frac{\partial \tilde{\Phi}}{\partial \tilde{t}}
&=& \left[-\tilde{\nabla}^2 + 2\tilde{V} +2\tilde{g}|\tilde{\Phi}|^2
+\right.\nonumber\\
&& \left. 2\int U_{\rm dd}(\tilde{\mathbf{x}} - \tilde{\mathbf{x'}})|\tilde{\Phi}
(\tilde{\mathbf{x'}})|^2d\tilde{\mathbf{x'}}\right] \tilde{\Phi},
\label{Eq.4}
\end{eqnarray}
where $\tilde V = (\lambda_x^2\tilde{x}^2+\lambda_y^2\tilde{y}^2
+\lambda_z^2\tilde{z}^2)/2$, $\lambda_j = \omega_j/\omega$
with $j = x,y,z$ and $\tilde g = 4\pi\hbar a N/(m\omega a_{\rm osc}^3)$.
In order to simplify the notations, from here on we will write these
variables without tildes unless mentioned
otherwise. In the present work, we consider the DBEC in a quasi-two-dimensional
trap for which $\lambda_x=\lambda_y=1\ll\lambda_z$. 
In this case, the axial
degrees of freedom of the system are frozen, and 
the chemical potential in scaled units $\mu_{3d}<\lambda_z$ \cite{Gorlitz}.
Here the chemical potential $\mu_{3d}$ in scaled units is defined as
\begin{eqnarray}
\mu_{3d} &=& \int \left[\frac{|\nabla \Phi(\mathbf{x})|^2}{2} 
+ V(\mathbf{x})|\Phi(\mathbf{x})|^2 +g|\Phi(\mathbf{x})|^4+
\right.\nonumber\\
&& \left. \Phi^*(\mathbf{x})\int U_{\rm dd}(\mathbf{x} - \mathbf{x'})|\Phi
(\mathbf{x'})|^2d\mathbf{x'}\right] \Phi(\mathbf{x}) d\mathbf{x}.
\label{chem_pot}
\end{eqnarray} 
We write
$\Phi(\mathbf x) = \zeta(z)\phi(x,y)$ with
$\zeta(z) = (\lambda_z/\pi)^{1/4}e^{-(\lambda_z z^2)/2}$ as the harmonic
oscillator ground state along the axial direction. After integrating out the
axial coordinate, we obtain the following two-dimensional equation
\cite{Pedri, Muruganandam-1,Fischer,Ticknor}:
\begin{eqnarray}
i\frac{\partial \phi}{\partial t} &=& \left\{-\nabla_{\rho}^2
+ 2V_{\rho} +2g_{\rho}|\phi|^2 +4\sqrt{2\pi\lambda_z}a_{\rm dd}N\right.
\nonumber\\
&& \times\int \frac{d^2k_{\rho}}{4\pi^2}
e^{i \mathbf{k}_{\rho}.\tilde{\rho}}\tilde{n}(\mathbf k_{\rho})
\left[\cos^2(\alpha)h_{2d}^{\perp}\left(\frac{\mathbf{k}_{\rho}}
{\sqrt{2\lambda_z}}\right) \right.\nonumber\\
&&\left.\left.+\sin^2(\alpha)h_{2d}^{\parallel}\left(\frac{\mathbf{k}_{\rho}}
{\sqrt{2\lambda_z}}\right)\right]\right\} \phi,
\label{gpe_scaled}
\end{eqnarray}
where $\nabla_{\rho}^2 = \partial^2/\partial x^2+ \partial^2/\partial y^2$,
$V_{\rho} = x^2/2+y^2/2$ and $g_{\rho} = \sqrt{\lambda_z/2\pi}g$. It must
be pointed out here that the term $\lambda_z\phi$, arising from the axial
energy, has been neglected from the right hand side of the Eq.~(\ref{gpe_scaled}) 
as it only decreases the chemical potential and 
energy by an amount of $\lambda_z\hbar\omega/2$ without affecting the dynamics \cite{Salasnich}. 
Here we have considered an arbitrary direction of polarization in the $xz$ plane, which
makes an angle $\alpha$ with the $z$ axis, and
$h_{2d}^{\perp}(\mathbf{k}_{\rho}/\sqrt{2\lambda_z})$ and
$h_{2d}^{\parallel}(\mathbf{k}_{\rho}/\sqrt{2\lambda_z})$ are defined as
\begin{eqnarray}
h_{2d}^{\perp}\left(\frac{\mathbf{k}_{\rho}}{\sqrt{2\lambda_z}}\right) & =
&2-\frac{3\sqrt{\pi}k_{\rho}}{\sqrt{2\lambda_z}}
\exp\left(\frac{k_{\rho}^2}{2\lambda_z}\right){\rm erfc}\left(\frac{k_{\rho}}
{\sqrt{2\lambda_z}}\right),\nonumber\\
h_{2d}^{\parallel}\left(\frac{\mathbf{k}_{\rho}}{\sqrt{2\lambda_z}}\right) & =
& -1+\frac{3\sqrt{\pi}k_x^2}{\sqrt{2\lambda_z}k_{\rho}} \exp\left(\frac{k_{\rho}^2}{2\lambda_z}
\right){\rm erfc}\left(\frac{k_{\rho}}{\sqrt{2\lambda_z}}\right).\nonumber
\end{eqnarray}
The scaled wavefunction is normalized to unity, i.e., $\int|\phi|^2d\rho = 1$.
We use the time splitting Fourier spectral method to solve equation
Eq.~(\ref{gpe_scaled}) \cite{Bao}. The spatial and time step sizes employed in
the present work are $0.1~a_{\rm osc}$ and $0.0005~{\omega^{-1}}$
respectively.


\section{Merging and splitting dynamics of the DBEC}
\label{Sec-III}
In the present work, we consider $5\times10^4$ (or $10^5$) atoms of $^{52}$Cr 
in a trapping potential with $\omega_x = \omega_y = \omega = 2\pi\times10$ Hz 
and $\omega_z = 2\pi\times980$ Hz. Hence, the units of length and time employed 
are $a_{\rm osc} = 4.4~\mu$m and $\omega^{-1} = 1.59\times10^{-2}$s respectively. 
The dipolar length of $^{52}$Cr is $16~a_0$ and the background $s$-wave scattering 
length is $100~a_0$ \cite{Lahaye-1}. The $s$-wave scattering length of $^{52}$Cr 
can be tuned by magnetic Feshbach resonances \cite{Lahaye-1,Werner}. However, we 
consider $a = 8~a_0$, which is significantly smaller than its background value,
in order to accentuate the effects of dipolar interactions.

\begin{center}
\begin{figure}[!h]
\begin{tabular}{c}
\resizebox{!}{!}
{\includegraphics[trim = 0cm 0cm 0cm 0cm,clip, angle=0,width=4.0cm]
                 {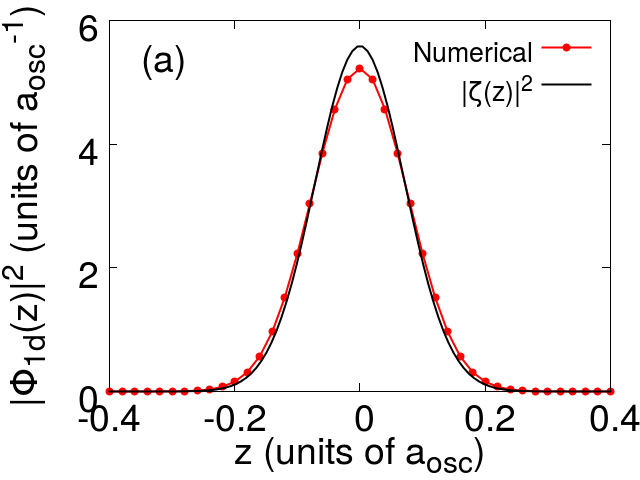}}
\resizebox{!}{!}
{\includegraphics[trim = 0cm 0cm 0cm 0cm,clip, angle=0,width=4.0cm]
                 {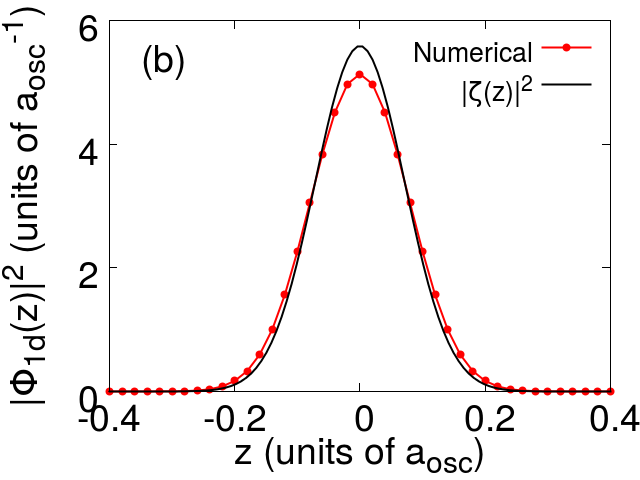}}
\end{tabular}
\caption{Variation of $|\Phi_{1d}(z)|^2$ corresponding to the ground
state solution and $|\zeta(z)|^2$ as a function
of $z$ for (a) $N = 5\times10^4$ and (b) $N = 10^5$. The black and red
curves correspond to $|\zeta(z)|^2$ and $|\Phi_{1d}(z)|^2$ respectively.}
\label{fig-1}
\end{figure}
\end{center}

Before proceeding further, let us first analyze the validity
of the Eq.~(\ref{gpe_scaled}) used in the present work. To this end,
we solve the full three dimensional GP equation (\ref{Eq.1}) with
$N = 5\times10^4$ and $N = 10^5$ atoms of the $^{52}$Cr for the aforementioned
trapping potential parameters with $a = 8~a_0$, $a_{\rm dd} = 16 a_0$, and 
with dipoles oriented along $z$-axis ($\alpha = 0$). We use imaginary time
propagation method, where $t$ is replaced by $-i\tau$, to obtain the ground state 
solution of Eq.~(\ref{Eq.1}). 
The spatial and temporal (imaginary time) step sizes used to solve 
Eq.~(\ref{Eq.1}) are $\delta x = \delta y = 0.1$, 
$\delta z= 0.02$, and $\delta \tau = 0.0005$ respectively.
By integrating out the $x$ and $y$ dependence of the total density $|\Phi(x,y,z)|^2$,
we calculate the one-dimensional density along $z$-axis, i.e.,
\begin{equation}
 |\Phi_{1d}(z)|^2 = \int |\Phi(x,y,z)|^2dx dy.
\end{equation}
In Fig.~\ref{fig-1}, we have shown the variation of this one-dimensional density
$|\Phi_{1d}(z)|^2$ corresponding to the ground state solution and $|\zeta(z)|^2$ with respect to axial coordinate.
As is evident from the Figs.~\ref{fig-1}(a) and (b), the one-dimensional density
along the axial direction is very close to the density corresponding to
the harmonic oscillator ground state along the same direction. This justifies
the splitting of total wavefunction as $\Phi(x,y,z) = \phi(x,y)\zeta(z)$, and
therefore the use of Eq.~(\ref{gpe_scaled}) in the present work. 
Moreover, the chemical potential values obtained by solving the full three-dimensional
GP Eq.~(\ref{Eq.1}) and then using Eq.~(\ref{chem_pot}) 
are $\mu_{3d} = 66.0~\hbar\omega$ and $\mu_{3d} = 73.1~\hbar\omega$  
for $N = 5\times10^4$ and $N = 10^5$ respectively. 
Hence, $\mu_{3d}< \hbar\omega_z = 98~\hbar\omega$ in both the cases as is necessary for the system to
be in quasi-two-dimensional regime \cite{Gorlitz}.
Here, it is pertinent to point out that by replacing $\zeta(z)$ with an {\em ansatz}
which is the superposition of zeroth and second harmonic oscillator
wave functions with variable width and relative amplitude, a more
accurate two-dimensional GP equation has been suggested by 
Wilson {\em et al.} \cite{Wilson-4}. It has also been shown in the context of
quasi-two-dimensional bright solitons in the dipolar condensates that the non-linear coupling
can lead to the deposition of the excitation energy along the tightly bound 
direction \cite{Eichler}. However, very strong confinement along axial direction
in the present work ensures that the frozen Gaussian approximation along
this direction is a reasonably good approximation.

\subsection{Merging of two DBEC fragments}
In order to divide the DBEC into two fragments, we apply a Gaussian 
obstacle potential, in addition to the harmonic trapping potential mentioned earlier,
on the condensate. Now, to study the effect of
anisotropic superfluidity, introduced by a non-zero component of the dipoles in the $xy$
plane, on the merging dynamics, we consider the following two obstacle
potentials
\begin{eqnarray}
V^{\perp}_{\rm obs} &=& V_0 \exp\left(-2\frac{x^2}{w_0^2}\right),
\label{obs_pot_x}\\
V^{\parallel}_{\rm obs} &=& V_0 \exp\left(-2\frac{y^2}{w_0^2}\right),
\label{obs_pot_x}
\end{eqnarray}
here $V_0 = 100~\hbar\omega$ is the amplitude of the Gaussian potential
and $w_0 = 2.5~\mu$m is its width. Applying $V^{\perp}_{\rm obs}$ creates a repulsive
barrier potential along $y$-axis, i.e., perpendicular to the tilt of the dipoles on
$xy$-plane, whereas applying $V^{\parallel}_{\rm obs}$ creates a
repulsive barrier potential along the $x$-axis, i.e., parallel to the tilt of the dipoles. 
We now consider two cases in order
to contrast the anisotropic merging dynamics with the isotropic one.

\begin{center}
\begin{figure}[!h]
\includegraphics[trim = 0cm 0mm 0cm 0mm,clip, angle=0,width=8cm]
                 {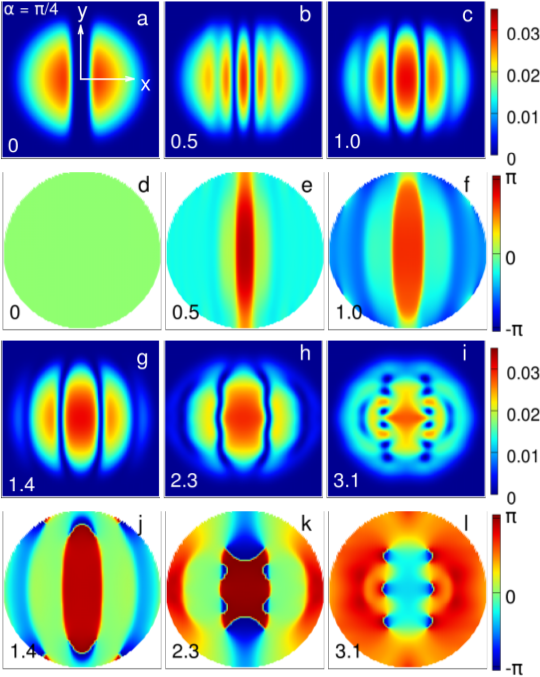}
\caption{Merging dynamics of two fragments of the DBEC along the $x$-axis
(directed along the horizontal direction in each image) for $\alpha = \pi/4$.
The obstacle potential $V^{\perp}_{\rm obs}$ is switched off at $t=0~\omega^{-1}$. 
The first and third rows show the densities of the DBEC, while the third and fourth 
rows show the corresponding phases at different times.  
The time in units of $\omega^{-1}$ is shown at the bottom left corner of each image.
The dimensions of each square image are $12~a_{\rm osc}\times12~a_{\rm osc}$,
and the origin is located at the center of each image.}
\label{fig-2}
\end{figure}
\end{center}

\begin{center}
\begin{figure}[!h]
\includegraphics[trim = 0cm 0mm 0cm 0mm,clip, angle=0,width=8cm]
                 {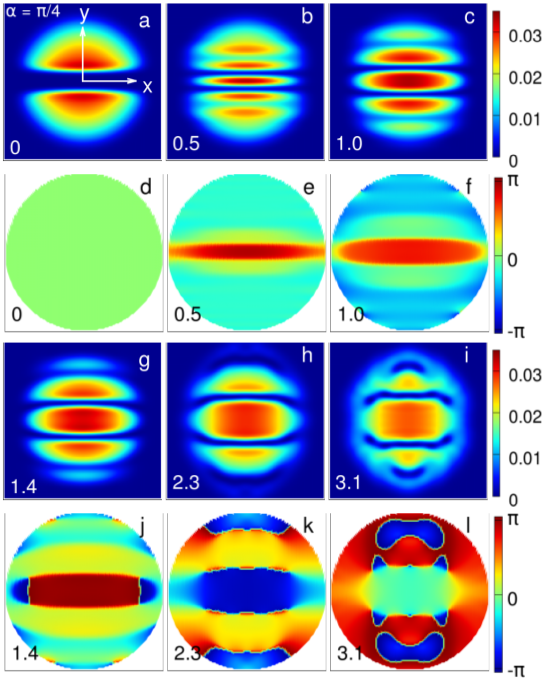}
\caption{Merging dynamics of two fragments of the DBEC along the $y$-axis
(directed along the vertical direction in each image) for $\alpha = \pi/4$.
The obstacle potential $V^{\parallel}_{\rm obs}$ is switched off at $t=0~\omega^{-1}$. 
The first and third rows show the densities of the DBEC, while the third and fourth 
rows show the corresponding phases at different times.  
The time in units of $\omega^{-1}$ is shown at the bottom left corner of each image.
The dimensions of each square image are $12~a_{\rm osc}\times12~a_{\rm osc}$,
and the origin is located at the center of each image.}
\label{fig-3}
\end{figure}
\end{center}

\begin{center}
\begin{figure}[!h]
\includegraphics[trim = 0cm 0mm 0cm 0mm,clip, angle=0,width=8cm]
                 {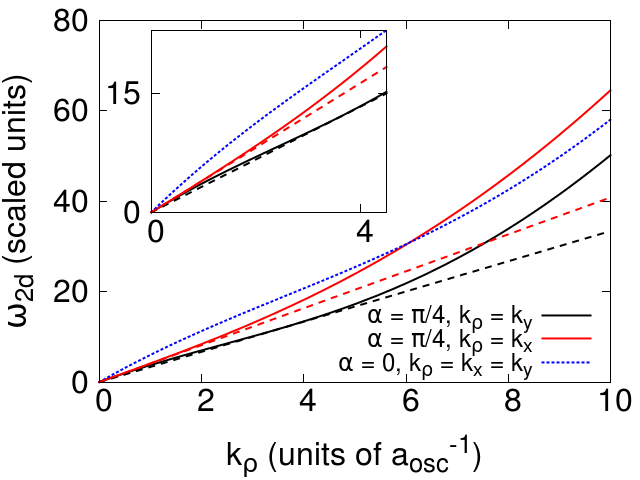}
\caption{The solid red and black lines are the Bogoliubov dispersion for
 the excitations propagating along the $x$- and $y$-axis respectively for
 $\alpha = \pi/4$. The dashed
 red line is the tangent to these dispersion curves; its slope is equal
 to the speed of sound. The slope of the dashed black curve is equal to the Landau
 critical velocity along the $y$-axis. The dotted blue is the isotropic Bogoliubov
 dispersion for $\alpha = 0$. For all the curves $n = 0.035$, $N = 5\times10^4$,
 and $\lambda_z = 98$. Inset shows the zoom-in of the main figure.}
\label{disp}
\end{figure}
\end{center}

\begin{center}
\begin{figure}[!h]
\includegraphics[trim = 0cm 0cm 0cm 0cm,clip, angle=0,width=8.0cm]
                 {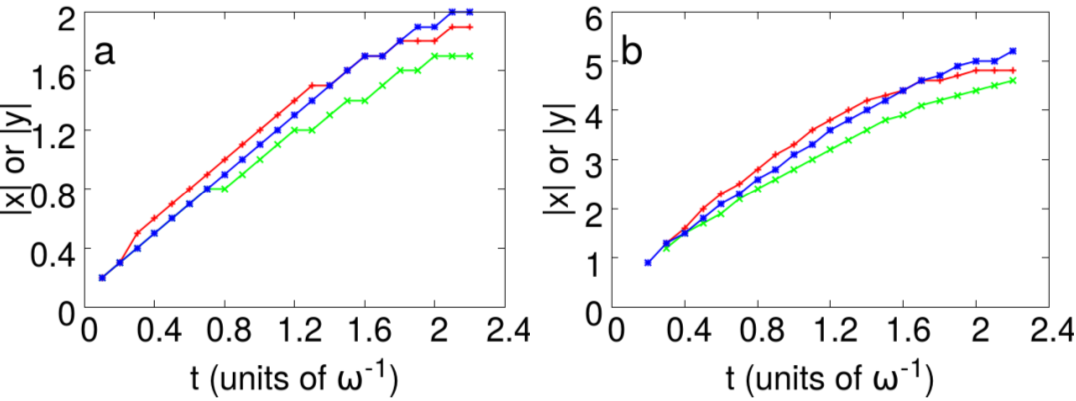}
\caption{The red (green) curve is the $x$ ($y$) coordinate of the mid-point of the 
 solitons shown in Fig.~\ref{fig-2} (Fig.~\ref{fig-3}). The blue curve is the
 $x$ coordinate of the mid-point of the solitons shown in 
 Fig.~\ref{iso_merg}. Sub-figure (a) corresponds to the two solitons near to the origin, 
 whereas the sub-figure (b) corresponds to the two solitons near the edge of the
 condensate.}
\label{fig-4}
\end{figure}
\end{center}

\begin{center}
\begin{figure}[!h]
\includegraphics[trim = 0cm 0mm 0cm 0mm,clip, angle=0,width=8cm]
                 {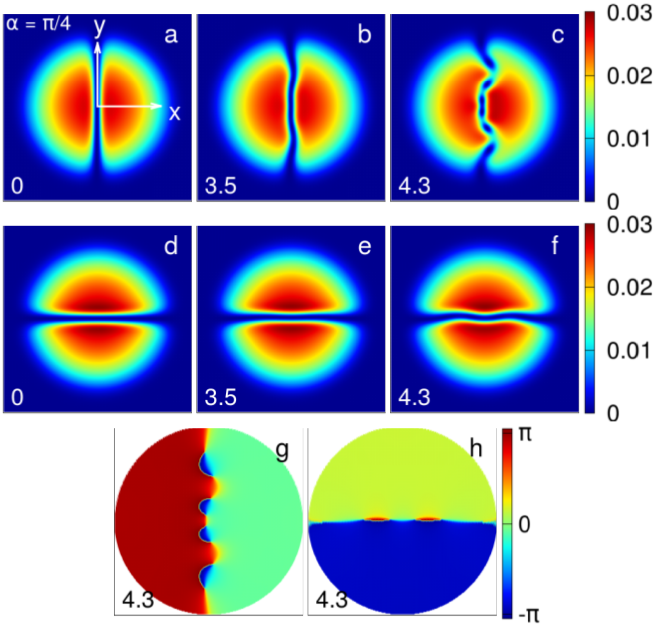}
\caption{The upper (lower) row shows the dynamics of the dark soliton imprinted along
 the $y$ ($x$) axis for a given period. The sub-figures (g) and (h) are
 the phase plots corresponding to density distributions shown in
 sub-figures (c) and (f). The time
 in units of $\omega^{-1}$ is shown at the bottom left corner of each image.
 The dimensions of each image are $12~a_{\rm osc}\times12~a_{\rm osc}$.}
\label{fig-5}
\end{figure}
\end{center}

\begin{center}
\begin{figure}[!h]
\includegraphics[trim = 0cm 0mm 0cm 0mm,clip, angle=0,width=8cm]
                 {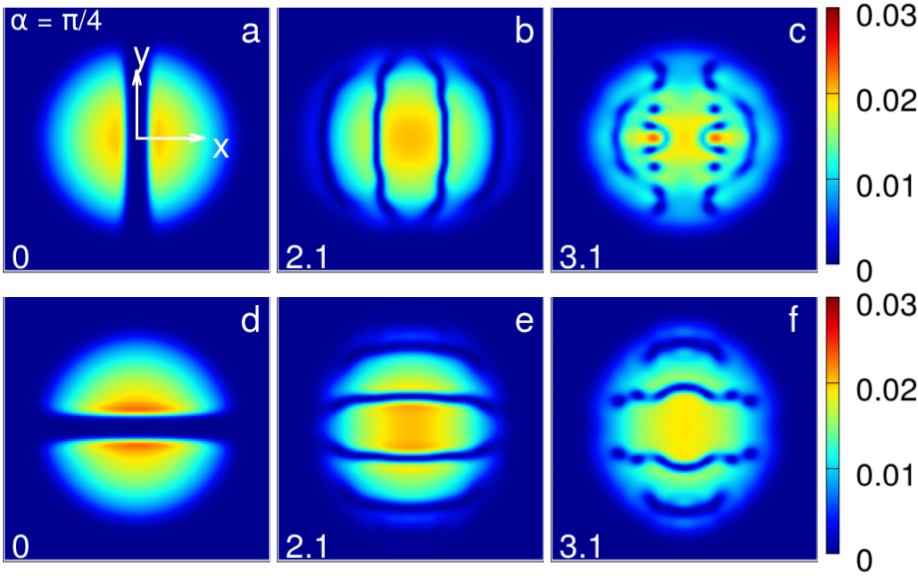}
\caption{Merging dynamics of the DBEC with $10^5$ atoms. The time
 in units of $\omega^{-1}$ is shown at the bottom left corner of each image.
 The dimensions of each image are $16~a_{\rm osc}\times16~a_{\rm osc}$.}
\label{merg_dyn_large}
\end{figure}
\end{center}

{\em Anisotropic merging:} Here we take the angle $\alpha$ to be $\pi/4$. We first obtain
the static solution by solving Eq.~(\ref{gpe_scaled}) using imaginary time
propagation for both obstacle potentials $V^{\perp}_{\rm obs}$ and $V^{\parallel}_{\rm obs}$.
The solutions thus obtained for the aforementioned two barrier potentials are
shown in Figs.~\ref{fig-2}(a) and (d) for $V^{\perp}_{\rm obs}$, and 
Figs.~\ref{fig-3}(a) and (d) for $V^{\parallel}_{\rm obs}$.
The energy of the DBEC without any obstacle potential is $8.11~\hbar\omega$.
This energy does not include $\hbar\omega_z/2$ contribution form the axial
direction which was neglected while writing Eq.~(\ref{gpe_scaled}) as
has been mentioned in the introduction. 
We find that energy cost of splitting the condensate along the $y$-axis is greater
than splitting it along the $x$-axis; the energy difference
$\Delta E = 0.24~\hbar\omega$ for the system studied in
Fig.~\ref{fig-2}(a) ($E = 9.96~\hbar\omega$) and Fig.~\ref{fig-3}(a) ($E = 9.72~\hbar\omega$). 
In case the dipoles are not oriented along
the $z$-axis, the dispersion relation has a directional dependence and for
a homogeneous two dimensional system is given by \cite{Fischer,Ticknor}
\begin{eqnarray}
\omega_{\rm 2d} (\mathbf{k}_{\rho})&=& \Bigg\{\frac{k_{\rho}^4}{4}+
2nNk_{\rho}^2\sqrt{2\pi\lambda_z}\bigg[a+a_{\rm dd}\Big(\cos^2(\alpha)
h^{\perp}_{\rm 2d} \nonumber\\
 & &+ \sin^2(\alpha) h^{\parallel}_{\rm 2d}\Big)\bigg]\Bigg\}^{1/2}.
\label{disp_rel}
\end{eqnarray}
The Bogoliubov dispersion obtained using this expression for $n = 0.035$
(peak density corresponding to images in Fig.~\ref{fig-2} and Fig.~\ref{fig-3}),
$N = 5.0\times10^4$, $\lambda_z = 98$, and $\alpha = \pi/4$ is shown in
Fig.~\ref{disp}.
Also, the speed of the sound
\begin{eqnarray}
c_{\rho} &=& \lim_{k_{\rho}\rightarrow 0} \frac{\omega_{\rm 2d}
             (\mathbf{k}_{\rho})}{k_{\rho}},\nonumber\\
         &=& \sqrt{2nN\sqrt{2\pi\lambda_z}[a+a_{\rm dd}(2\cos^2\alpha-\sin^2{\alpha})]},
\end{eqnarray}
and hence is isotropic. This is in contrast to three dimensional (homogeneous) 
DBEC where the speed of the sound \cite{Lahaye-2,Lima,Muruganandam} 
\begin{equation}
 c_{\rho} = \sqrt{\frac{nN}{m}\left[g+\frac{C_{\rm dd}}{3}(3\cos^2\Theta-1) \right]},
 \end{equation}
where $\Theta$ is the angle between the wavevector and the direction of polarization. 
Hence, the speed of the sound in three dimensional DBEC is anisotropic 
and it has been demonstrated experimentally \cite{Bismut-1}.
Now, applying $V^{\parallel}_{\rm obs}$ leads to large density variations along
the $y$-axis producing excitations that lie on the black curve in
Fig.~\ref{disp} including the roton-like mode. 
In the present work, we use the term roton-like mode to refer to the 
relative softening of the dispersion for the quasi-particle propagation perpendicular
to the tilt of the dipoles on the plane of the condensate as compared to the 
dispersion perpendicular to it \cite{Ticknor}. The relative softening at the intermediate momentum 
results in the inflection point on the dispersion curve (see the inset of Fig.~\ref{disp}) 
without leading to typical roton minimum \cite{Santos}, which can not arise in a 
quasi-two dimensional condensate \cite{PB_Blakie}.  
On the other hand, applying
$V^{\perp}_{\rm obs}$ leads to large density variations along the $x$-axis which
can excite only modes with energy greater than the roton-like mode as is evident
from the red curve in Fig.~\ref{disp}. Thus, the anisotropic response of the
condensate to the perturbing potential is another manifestation of the
anisotropic dispersion and roton-like mode in the excitation spectrum, which
makes the quasi-particle excitations along the $y$-axis cost less energy.
Now, in order to allow the two fragments of the DBEC to merge non-adiabatically,
we suddenly switch off the obstacle potentials. This leads to the generation
of a train of dark notches. We identify these dark notches as solitons. The 
reasons for this identification are as follows. The
dark solitons in quasi-two-dimensional condensates exhibit two competing instability 
mechanisms \cite{Carr}. Firstly, a long wavelength sinusoidal mode transverse to the 
soliton grows exponentially, 
deforming the dark soliton into a snake like form. Later on, the arcs of this `snake'
like soliton decay into vortex-antivortex pairs. This instability, known as snake
instability \cite{Kadomtsev,Jones}, has nothing to with the presence of the trapping potential and even
occurs in the homogeneous two- and three-dimensional condensates. Secondly, a dark soliton
propagates at the fraction of the speed of the sound which depends on its depth,
and the speed of the sound is directly proportional to the square root of the
density of the condensates \cite{Dalfovo}. Hence, in the presence of the trapping potential, the inhomogeneous 
density profile causes the soliton to travel more slowly at the edges of the condensate 
than the center. As a result of it, an initially straight soliton
formed near the center of the trap deforms into a curved shape \cite{Carr}.
We find that during the initial stages of the merging dynamics, the solitons
become curved due to the inhomogeneity driven instability as is evident from
Figs.~\ref{fig-2}(b), (c), (g) and Figs.~\ref{fig-3}(b), (c), (g). The increase
in the curvature of the solitons is even more discernible in the phase plots
shown in Figs.~\ref{fig-2}(e), (f), (j) and Figs.~\ref{fig-3}(e), (f), (j).
After some time has elapsed, due to the snake instability the pair of
the solitons near the trap center deforms into a sinusoidal shape as
is evident from the Fig.~\ref{fig-2}(h) and Fig.~\ref{fig-3}(i).
Moreover, the long lifetime, large amplitude, and phase structure of these
dark notches also suggest their solitonic character. Our identification of
these dark notches as the solitons is consistent with the experimental studies 
\cite{Dutton, Chang, Hau}. These dark solitons are thus qualitatively different
from the bright solitons in quasi-two dimensional DBECs which are stable and can
move with a constant speed maintaining their shape \cite{Pedri, Tikhonenkov, R_Eichler}.
Now, the soliton train consists of four clearly 
discernible dark (grey) solitons. This is evident from Figs.~\ref{fig-2}(b), (c), (e), (f) 
and  Figs.~\ref{fig-3}(b), (c), (e), and (f).
We find that the separation between the adjacent solitons is not the same
in the two cases considered. This is
due to the fact that the solitons travel faster along the direction of
polarization, i.e. the $x$-axis as compared to the $y$-axis. In order to clearly
demonstrate this, we have measured the $x$ ($y$) coordinates of the mid-points of
the solitons. The variations of these coordinates are
shown by the red and green curves in Fig.~{\ref{fig-4}}. It is evident from
this figure that solitons propagating along the $x$-axis travel faster than
the ones propagating along the $y$-axis.
We also observe
that the solitons oriented along the direction of polarization are less
susceptible to the snake instability as compared to the ones oriented
perpendicular to it. This becomes clear by comparing Fig.~\ref{fig-2}(i), 
where the soliton pair near the trap center has already decayed into 
vortex-antivortex pairs, with Fig.~\ref{fig-3}(i), where the pair has
only acquired a sinusoidal shape.
This is due to the fact that the sinusoidal perturbation
on the soliton oriented along the polarization direction costs more energy
due to the anisotropy introduced by non zero value of $\alpha$. To confirm
this inference, we have also numerically studied the dynamics of the DBEC which
has a single dark soliton either along the $x$-axis or the $y$-axis. To do so,
we first generate the soliton by imprinting a phase jump of $\pi$ across the
$x$ or $y$-axis. The static solution thus obtained is shown in the first column
of Fig.~\ref{fig-5}.
Again, as in the case of the obstacle potential,
imprinting a soliton along the $x$-axis costs less energy due to the excitation
of the roton like mode. We then evolve this solution in real time and find that
by the time the soliton oriented along the $y$-axis acquires a sinusoidal
shape due to snake instability, there is no perceptible change in the structure of the
soliton oriented along the $x$ axis (see the middle column of Fig.~\ref{fig-5}).
Later on at $4.3~\omega^{-1}$, when the soliton oriented along the $y$ axis has split 
into vortex-antivortex pairs due to the snake instability, the other soliton 
has merely acquired the sinusoidal shape due the snake instability as is shown in the 
last column of Fig.~\ref{fig-5}. The dipole-dipole interactions are also known to stabilize the three-dimensional
dark solitons against the snake instability in the presence of an optical lattice in
the nodal plane \cite{Nath}.
We have also studied the merging dynamics of a much larger condensate with
$N = 10^5$ with the rest of the parameters remaining unchanged.
We find that for the larger condensate, the difference in the merging
dynamics along the two orthogonal directions becomes more pronounced as is
shown in Fig.~\ref{merg_dyn_large}, and hence can be easily verified in 
current experiments.
This is due to the fact that by increasing $N$ while keeping the trapping
potential parameters unchanged, density ($N|\phi|^2 = Nn$) increases. This
results in increasing the difference in the slopes of the two dashed curves
in Fig.~\ref{disp}. In other words, the energy of the roton like mode is
lowered, while the energy cost of exciting modes along $x$-axis increases. It
implies that the anisotropy in the superfluidity of the system increases
with the increase in the number of atoms. This leads to the significantly
different merging dynamics along the $x$ and $y$-axes.

\begin{center}
\begin{figure}[!h]
\includegraphics[trim = 0cm 0mm 0cm 0mm,clip, angle=0,width=8cm]
                 {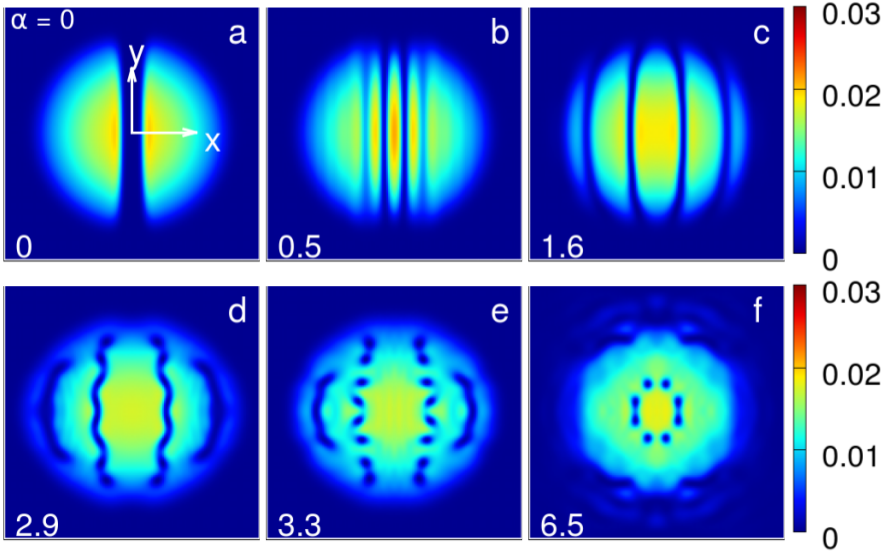}
\caption{The merging dynamics of two
fragments of dipolar condensates along the $x$ axis for $\alpha=0$.
The time in units of $\omega^{-1}$ is shown at the bottom left corner of each image.
 The dimensions of each image are $16~a_{\rm osc}\times16~a_{\rm osc}$.}
\label{iso_merg}
\end{figure}
\end{center}

{\em Isotropic merging:} Here we consider $\alpha = 0$, i.e. the dipoles are
oriented along the $z$ axis and $N = 5\times10^4$. The trapping and obstacle
potentials are the same as those considered in anisotropic merging dynamics.
Again, we first obtain the static solution by solving
Eq.~(\ref{gpe_scaled}) using imaginary time propagation for both obstacle
potentials $V^{\perp}_{\rm obs}$ and $V^{\parallel}_{\rm obs}$. The solution thus obtained
for the aforementioned barrier potential $V^{\perp}_{\rm obs}$ is shown in
Fig.~\ref{iso_merg}(a).
We then suddenly switch off the obstacle potentials. This leads to the
generation of a soliton train as in the previous case, which again consists of
four clearly discernible dark (grey) solitons. This is evident from the
Figs.~\ref{iso_merg}(b), (c). We observe the exactly identical dynamics by
using $V^{\parallel}_{\rm obs}$ instead of $V^{\perp}_{\rm obs}$.
This is due to fact that the Bogoliubov quasi-particle
dispersion obtained from Eq.~\ref{disp_rel} is isotropic for $\alpha = 0$ as 
is shown by the dotted blue line in in Fig.~\ref{disp} for $ n = 0.035$
and $N = 5\times10^4$. It is evident that the dispersion relation still has
roton-like features, but no directional dependence. In other words,
the roton like mode in this case is isotropic and hence the
dynamics of merging is independent of direction.

\subsection{Anisotropic splitting of the DBEC}
\begin{center}
\begin{figure}[!h]
\includegraphics[trim = 0cm 0mm 0cm 0mm,clip, angle=0,width=8cm]
                 {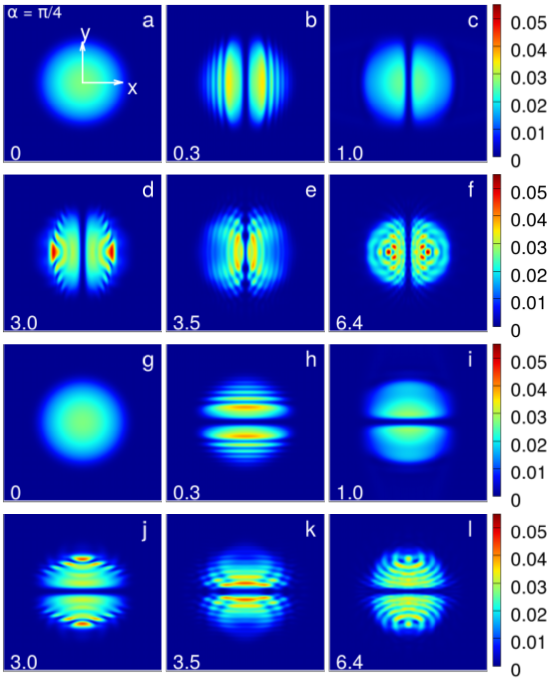}
\caption{Anisotropic splitting dynamics of the DBEC. The time
 in units of $\omega^{-1}$ is shown at the bottom left corner of each image.
 The dimensions of each image are $16~a_{\rm osc}\times16~a_{\rm osc}$.}
\label{aniso_spl}
\end{figure}
\end{center}

In order to study the anisotropic splitting of the DBEC, we consider
$5\times10^4$ atoms of $^{52}$Cr. In order to avoid the reflection of the
DBEC from the edges of the grid, we consider
a sufficiently large grid.
The grid spacing, trapping potential and the scattering length values are
same as those considered in the previous subsection. Here we first generate
the static solution without any obstacle potential and consider
$\alpha = \pi/4$. The static solution thus obtained is shown in
Figs.~\ref{aniso_spl}(a) and (g).
We then suddenly introduce
the obstacle potential either along the $x$ or $y$-axis. The strength and
the width of the obstacle potentials are $50~\hbar\omega$ and $1~\mu$m
respectively. The sudden turn-on of the obstacle potential produces
sharp density gradients with several density peaks as is shown in
Figs.~\ref{aniso_spl}(b) and (h), a possible signature of shock waves
\cite{Hoefer}. After some time, the broadest of these
density peaks slowly grows at the expense of the others.
During this period, the formation of the dispersive shock wave also leads
to the spilling of some of the condensate atoms beyond the edge of the
condensate (see Figs.~\ref{aniso_spl}(c) and (i)). We observe that during
the initial stages of the evolution, the dynamics of splitting along the
two orthogonal directions is almost identical as is evidenced by the
comparison of Figs.\ref{aniso_spl}(b) and (c) with
Figs.~\ref{aniso_spl}(h) and (i), respectively. The effect of anisotropy
starts manifesting itself during the latter stages of evolution as is
evidenced by a difference in the density distributions shown in
Figs.~\ref{aniso_spl}(d), (e), and (f) vis-\'a-vis Figs.~\ref{aniso_spl}(j),
(k), and (f). Hence, the dynamics of the splitting depends upon the
direction along which the barrier is introduced. We also find that increasing
the number of atoms to $10^5$ does not lead to any qualitative
differences in the splitting dynamics. This may make it difficult to 
observe anisotropic splitting in current experiments. Also, the anisotropy 
in dynamics disappears for $\alpha = 0$ as was the case for the merging 
dynamics.

\section{Summary of results}
\label{Sec-IV}
We have numerically studied the dynamics of non-adiabatic merging and
splitting of the dipolar Bose-Einstein condensate. The non-adiabatic merging
and splitting is achieved by suddenly removing or applying an obstacle
potential on the condensate. For the sake of observing the signature of
anisotropic superfluidity, we implement the merging and splitting of the
condensate along
two orthogonal directions, one of which is parallel to the dipoles' projection on
the plane of the condensate. We observe that if the direction of polarization
is not normal to the plane of the condensate, the roton-like features of
the dispersion are manifested by the directional dependence of merging and
splitting dynamics. The absence of the anisotropy in the merging and splitting
dynamics rules out the existence of the anisotropic roton-like mode, although
the isotropic roton like mode can still exist. From the experimental perspective,
the tunability of the Bogoliubov dispersion by changing the density can be used to
either increase or decrease the effects of anisotropic superfluidity on the
dynamics of the DBEC. Our studies
indicate that although anisotropic splitting may be difficult to
observe experimentally, there should be no such issue with
anisotropic merging dynamics.


\begin{acknowledgements}
The authors would like to thank the Department of Science and Technology,
Government of India for support.
\end{acknowledgements}



\begin{thebibliography}{99}
\bibitem{Griesmaier}
 A.~Griesmaier, J.~Werner, S.~Hensler, J.~Stuhler, and T.~Pfau,
 Phys. Rev. Lett. {\bf 94}, 160401 (2005);
 A.~Griesmaier, J.~Stuhler, and T.~Pfau,
 Applied Physics B {\bf 82}, 211 (2006).
\bibitem{Lahaye-1}
 T.~Lahaye, T.~Koch, B.~Fr\"ohlich, M.~Fattori, J.~Metz, A.~Griesmaier,
 S.~Giovanazzi, and T.~Pfau,
 Nature {\bf 448}, 672 (2007).
\bibitem{Koch}
 T.~Koch, T.~Lahaye, J.~Metz, B.~Fr\"ohlich, A.~Griesmaier, and T.~Pfau,
 Nature Physics {\bf 4}, 218 (2008).
\bibitem{Lahaye-2}
 T.~Lahaye, C.~Menotti, L.~Santos, M.~Lewenstein, and T.~Pfau,
 Rep. Prog. Phys. {\bf 72}, 126401 (2009).
\bibitem{Baranov}
 M.~A.~Baranov, M.~Dalmonte, G.~Pupillo, and P.~Zoller,
 Chem. Rev. {\bf 112}, 5012 (2012).
\bibitem{Lu-1}
 M.~Lu, N.~Q.~Burdick, S.~H.~Youn, and B.~L.~Lev,
 Phys. Rev. Lett. {\bf 107}, 190401 (2011),
\bibitem{Aikawa}
 K.~Aikawa, A.~Frisch, M.~Mark, S.~Baier, A.~Rietzler, R.~Grimm,
 and F.~Ferlaino,
 Phys. Rev. Lett. {\bf 108}, 210401 (2012).
\bibitem{Lu-2}
 M.~Lu, N.~Q.~Burdick, and B.~L.~Lev,
 Phys. Rev. Lett. {\bf 108}, 215301 (2012).
\bibitem{K_Aikawa}
 K.~Aikawa, A.~Frisch, M.~Mark, S.~Baier, R.~Grimm, and F.~Ferlaino,
 Phys. Rev. Lett. {\bf 112}, 010404 (2014).
\bibitem{Stuhler}
 J.~Stuhler, A.~Griesmaier, T.~Koch, M.~Fattori, T.~Pfau,
 S.~Giovanazzi, P.~Pedri, and L.~Santos, 
 Phys. Rev. Lett. {\bf 95}, 150 406 (2005).
\bibitem{Bismut-2}
 G.~Bismut, B.~Pasquiou, E. Mare\'chal, P.~Pedri, L.~Vernac,
 O.~Gorceix, and B.~Laburthe-Tolra, 
 Phys. Rev. Lett. {\bf 105}, 040404 (2010).
\bibitem{Santos}
 L.~Santos, G.~V.~Shlyapnikov, M.~Lewenstein,
 Phys. Rev. Lett. {\bf 90}, 250403 (2003).
\bibitem{Wilson-1}
 R.~M.~Wilson, S.~Ronen, J.~L.~Bohn, and H.~Pu,
 Phys. Rev. Lett. {\bf 100}, 245302 (2008).
\bibitem{Wilson-2}
 R.~M.~Wilson, S.~Ronen, and J.~L.~Bohn,
 Phys. Rev. Lett. {\bf 104}, 094501 (2010).
\bibitem{Landau}
 L.~Landau,
 J. Phys. (Moscow) {\bf 5}, 71 (1941).
\bibitem{Yi}
 S.~Yi and H.~Pu,
 Phys. Rev. A {\bf 73}, 061602(R) (2006).
\bibitem{Ronen}
 S.~Ronen, D.~C.~E.~Bortolotti, and J.~L.~Bohn,
 Phys. Rev. Lett. {\bf 98}, 030406 (2007).
\bibitem{Wilson-3}
 R~M.~Wilson, S.~Ronen, and J.~L.~Bohn,
 Phys. Rev. A {\bf 80}, 023614 (2009).
\bibitem{Blakie}
 P.~B.~Blakie, D.~Baillie, and R.~N.~Bisset,
 Phys. Rev. A {\bf 86}, 021604(R) (2012.
\bibitem{Bisset}
 R.~N.~Bisset and P.~B.~Blakie,
 Phys. Rev. Let. {\bf 110}, 265302 (2013).
\bibitem{Corson-1}
 J.~P.~Corson, R.~M.~Wilson, and J.~L.~Bohn,
 Phys. Rev. A {\bf 87}, 051605 (2013).
\bibitem{Corson-2}
 J.~P.~Corson, R.~M.~Wilson, and J.~L.~Bohn,
 Phys. Rev. A {\bf 88}, 013614 (2013).
\bibitem{Jona-Lasinio-1}
 M.~Jona-Lasinio, K.~Lakomy, and L.~Santos,
 Phys. Rev. A {\bf 88}, 025603 (2013).
\bibitem{Boudjemaa}
 A.~Boudjem\^{a}a and G.~V.~Shlyapnikov,
 Phys. Rev. A {\bf 87}, 025601 (2013).
\bibitem{Jona-Lasinio-2}
 M.~Jona-Lasinio, K.~Lakomy, and L.~Santos,
 Phys. Rev. A {\bf 88}, 013619 (2013).
\bibitem{Fischer}
 U.~R.~Fischer,
 Phys. Rev. A {\bf 73}, 031602(R) (2006).
\bibitem{Ticknor}
 C.~Ticknor, R.~M.~Wilson, and J.~L.~Bohn,
 Phys. Rev. Lett. {\bf 106}, 065301 (2011).

\bibitem{Muruganandam}
 P.~Muruganandam and S.~K.~Adhikari,
 Phys. Lett. A {376}, 480 (2012).
\bibitem{Bismut-1}
 G. Bismut, B. Laburthe-Tolra, E. Mare\'chal P. Pedri, O. Gorceix, and L. Vernac,
 Phys. Rev. Let. {\bf 109}, 155302 (2012).
 \bibitem{Ticknor-2}
 C.~Ticknor, 
 Phys. Rev. A {\bf 86}, 053602 (2012).
\bibitem{Chang}
 J.~J.~Chang, P.~Engels, and M.~A.~Hoefer,
 Phys. Rev. Lett. {\bf 101}, 170404 (2008).
\bibitem{Dutton}
 Z.~Dutton, M.~Budde, C.~Slowe, and L.~V.~Hau,
 Science {\bf 293}, 663 (2001).
\bibitem{Joseph}
 J.~A.~Joseph, J.~E.~Thomas, M.~Kulkarni, and A.~G.~Abanov,
 Phys. Rev. Lett. {\bf 106}, 150401 (2011).
\bibitem{Bulgac}
 A.~Bulgac, Y-L Luo, and K.~J. Roche,
 Phys. Rev. Lett. {\bf 108} 150401 (2012).
\bibitem{Ancilotto}
 F.~Ancilotto, L.~Salasnich, and F.~Toigo,
 Phys. Rev. A, {\bf 85}, 063612 (2012).
\bibitem{Wan}
 W.~Wan , S.~Jia, and J.~W.~Fleischer,
 Nature Physics {\bf 3}, 46 (2006).
 \bibitem{Gorlitz}
 A.~G\"orlitz, J.~M.~Vogels, A.~E.~Leanhardt, C.~Raman, T.~L.~Gustavson, J.~R.~Abo-Shaeer,
 A.~P.~Chikkatur, S.~Gupta, S.~Inouye, T.~Rosenband, and W.~Ketterle,
 Phys. Rev. Lett. {\bf 87}, 130402 (2001).

\bibitem{Pedri}
 P.~Pedri and L.~Santos,
 Phys. Rev. Lett. {\bf 95}, 200404 (2005).
\bibitem{Muruganandam-1}
 P.~Muruganandam and S.~K.~Adhikari,
 Laser Physics {\bf 22}, 813 (2012).
 \bibitem{Salasnich}
 L.~Salasnich, A.~Parola, and L.~Reatto,
 Phys. Rev. A {\bf 65}, 043614 (2002).
\bibitem{Bao}
 W.~Bao, D.~Jaksch, and P.~A.~Markowich,
 J. Comp. Phys. {\bf 187}, 318 (2003).
\bibitem{Werner}
 J.~Werner, A.~Griesmaier, S.~Hensler, J.~Stuhler, T.~Pfau,
 A.~Simoni, and E.~Tiesinga
 Phys. Rev. Lett. {\bf 94}, 183201 (2005).
  \bibitem{Wilson-4}
 R.~M.~Wilson and J.~L.~Bohn
 Phys. Rev. A {\bf 83}, 023623 (2011).
 \bibitem{Eichler}
 R.~Eichler, J.~Main, and G.~Wunner,
 Phys. Rev. A {\bf 83}, 053604 (2011).
 \bibitem{Lima}
 A.~R.~P.~Lima and A.~Pelster,
 Phys. Rev. A {\bf 86}, 063609 (2012).
 \bibitem{PB_Blakie}
 P.~B.~Blakie, D.~Baillie, and R.~N.~Bisset,
 Phys. Rev. A {\bf 88}, 013638 (2013).
 \bibitem{Carr}
 {\em Emergent Nonlinear Phenomena in Bose-Einstein Condensates: 
 Theory and Experiment}, 
 edited by P. G. Kevrekidis, D. J. Frantzeskakis, and R. Carretero-Gonzalez 
 (Springer-Verlag, Berlin, 2009).
\bibitem{Kadomtsev}
 B.~B.~Kadomtsev, V.~I.~Petviashvili,
 Sov. Phys. Dokl. {\bf 15}, 539 (1970).
\bibitem{Jones}
 C.~A.~Jones, S.~J.~Putterman, P.~H.~Roberts,
 J. Phys. A {\bf 19}, 2991 (1986).
 \bibitem{Dalfovo}
 F.~Dalfovo, S.~Giorgini, L.~P.~Pitaevskii, and S.~Stringari, 
 Rev. Mod. Phys. {\bf 71}, 463 (1999).
 \bibitem{Hau}
 L.~V~Hau,
 Nature Physics {\bf 3}, 13 (2007).
\bibitem{Tikhonenkov}
 I.~Tikhonenkov, B.~A.~Malomed, and A.~Vardi, 
 Phys. Rev. Lett. {\bf 100}, 090406 (2008).
 \bibitem{R_Eichler}
 R.~Eichler, D.~Zajec, P.~K\"oberle, J.~Main, and G.~Wunner,
 Phys. Rev. A {\bf 86}, 053611 (2012).
\bibitem{Nath}
 R.~Nath, P.~Pedri, and L.~Santos,
 Phys. Rev. Lett. {\bf 101}, 210402 (2008).
\bibitem{Hoefer}
 M.~A.~Hoefer, M.~J.~Ablowitz, I.~Coddington, E.~A.~Cornell,
 P.~Engels, and V.~Schweikhard,
 Phys. Rev. A {\bf 74}, 023623 (2006).
\end{thebibliography}
\end{document}